\documentclass[a4paper,twoside]{article}

\usepackage{epsfig}
\usepackage{subcaption}
\usepackage{calc}
\usepackage{amssymb}
\usepackage{amstext}
\usepackage{amsmath}
\usepackage{amsthm}
\usepackage{multicol}
\usepackage{pslatex}
\usepackage{apalike}
\let\bibhang\relax
\usepackage{natbib}

\usepackage{lineno}

\usepackage[table,xcdraw]{xcolor}
\usepackage[colorlinks=false,urlcolor=black,hidelinks]{hyperref}

\usepackage{SCITEPRESS}     

\begin{document}

\title{User Interface Factors of Mobile UX: A Study with an Incident Reporting Application}

\author{\authorname{Lasse Einfeldt\sup{1}, Auriol Degbelo\sup{2}\orcidAuthor{0000-0001-5087-8776}
}
\affiliation{\sup{1}xdot GmbH, Münster, Germany}
\affiliation{\sup{2}Institute for Geoinformatics, University of Münster, Münster, Germany}
\email{lasseeinfeldt@gmail.com, auriol.degbelo@uni-muenster.de}
}

\keywords{mobile UX, mobile form design, map UX, environmental monitoring}

\abstract{ Smartphones are now ubiquitous, yet our understanding of user interface factors that maximize mobile user experience (UX), is still limited. This work presents a controlled experiment, which investigated factors that affect the usability and UX of a mobile incident reporting app. The results indicate that sequence of user interface elements matters while striving to increase UX, and that there is no difference between tab and scrolling as navigation modalities in short forms. These findings can serve as building blocks for empirically-derived guidelines for mobile incident reporting.}

\onecolumn \maketitle \normalsize \setcounter{footnote}{0} \vfill

\section{Introduction}
User experience (UX) has gained attention from many sides since the turn of the millennium. Although many authors mentioned the lack of clear definitions of user experience and no clear understanding of it in early research \citep{Forlizzi2000,mcarty2004,Hassenzahl2005,wright2005}, there has been a lot of research dealing with user experience since then. There has been reviews of UX in human-computer interaction (e.g. \cite{Bargas-Avila2011,Pettersson2018,Kieffer2019}) and work investigating factors of mobile UX in general (e.g. \cite{wigelius2009dimensions,arhippainen2003empirical}). Nonetheless, our understanding of \textit{user interface factors} which maximize mobile user experience is still limited. Investigating these factors is important for at least two reasons. First, mobile devices are peculiar by virtue of their size, their (still) relatively reduced processing power, and input modalities (e.g. post-WIMP interaction). That is, insights gathered while assessing user experience on desktop (or other devices) are not readily portable to the world of mobile UX. Second, an understanding of these factors is important to develop design heuristics, which can then be integrated as constraints during computer-generated user interface design \citep{Oulasvirta2017}.

This work is primarily concerned with the impact of positioning and type of navigation modality on the overall usability and user experience of mobile geospatial applications. Incident reporting apps (of which maps are a central component) are of interest here for two reasons. From the theoretical point of view (and as mentioned in \cite{Kray2017,roth2013interactive}), there are currently no consolidated set of guidelines that have emerged as to how to design interactions with maps. A few works \citep{schoning2014informing,arhippainen13tutorial,Kraak2018} made some useful recommendations, but these do not address the UX of mobile geospatial apps (i.e. maps) directly. Thus, research in this area must remain active for study outcomes to crystallize into empirically-derived guidelines in the future. From the practical point of view, maximizing UX in the context of incident reporting is a catalyst for uptake and prolonged use. Put differently, a positive user experience is crucial to guarantee prolonged contributions by
citizens. As a starting point for the work, the mobile app “Meine Umwelt” was used. The rationale for its choice is introduced next.

Reports of ecological data are an important data source for German federal agencies dealing with the preservation of the environment. Reported data can have several topics including findings of neophytes, illegally disposed waste, or endangered animals. This information can be used to find items of interest in the wild easily, instead of searching for them. A system used for reporting this data is the mobile app “Meine Umwelt” \citep{MeineUmwelt}. “Meine Umwelt” was created by the “Kooperation Umweltportale”, a cooperation of German federal states for the development of environment-related apps. More than 10.000 downloads in the Google Play Store, more than 3000 active users on Android, in addition to more than 2.250 unique downloads on iOS  build the current user base.  Since the app is developed for federal agencies in Germany, means for measuring the interactions (e.g. interaction logs) with the application cannot be implemented into the live app in the app stores. An assessment of the interaction with the app to identify possibilities for improvement can only happen through a lab-based study offline. 

The contributions of this work are twofold: (i) an investigation of the effect of map placements on the user experience within a mobile app for reporting ecological data: a takeaway from the experiment is that sequence of UI elements matters while trying to maximize mobile UX; (ii) a comparison of tab-based and scrolling navigation for short forms on mobile devices: contrary to \cite{Harms2015} who found that scrolling performs worse than tab for the interaction with long forms on smartphones, the difference between both interaction modalities for short forms was not significant in this work.

\section{Related Work}
As indicated in \citep{Ricker2018}, mobile devices enable users to volunteer their local knowledge and experience while situated in place, providing timely and unstructured information about changing geographic conditions. Since “Meine Umwelt” is dedicated to the reporting of environmental data, related work on reporting systems is briefly presented in this section. In addition, research on factors of mobile UX is briefly introduced, followed by work on form design on mobile devices, and interactive maps.

\subsubsection*{Reporting systems}
\cite{winkler2016} focused on the UX dimensions important for incident reporting systems. Interviews with participants were used to gain insights into the users’ perception of the investigated app. They found users to prefer a selection of reportable items from a menu. This provides an overview of all items and avoids generic forms trying to fit all items at once. Moreover, an interactive map was clearly demanded by the users. Another requirement found in their interviews was the necessity to provide an identification, to avoid fake reports. Pictures and videos of the item should also be included in the report. Related to incident reporting are citizen observatories, defined as “the participation of citizens in monitoring the quality of the environment they live in” \citep{liu2014conceptual}. Here, no recent incident or event is present as a motivation for the report. Instead, observatories are more focused on ecological data and observations than on incidents in an urban context. The acquired data should be used by the government and if possible made available to the public as data and/or service \citep{liu2014conceptual}. Citizens’ observatories are described as a cheap and easy way for the administration to collect data about the environment.

The collection of data about the environment, where the citizens live is exactly what the prototypes presented later in this study were designed for. In addition, targeting a broad audience alongside the environmentalists, who are highly active in preservation anyway, is possible with a mobile application. Having data reported by local citizens as stakeholders can lead to enhanced quality in environmental decisions \citep{reed2008stakeholder}. This data is easy to collect and as a review of ten years of community-based monitoring found, one of the most efficient methods of monitoring \citep{conrad2011review}.  
\cite{Preece2016} references multiple technologies and concepts used for citizen science, like passive and active data collection, mobile apps, web portals, webcams, drones, and gamification. All of these can be useful approaches to collect data, which fulfill some given requirements. In addition, these concepts can be combined - like an active data collection platform available both as an app and web portal. She also mentions some reporting systems like Floracaching, iSpotNature.org, and iNaturalist.org. The study in this work uses active data collection via mobile apps. The domains of the reports are comparable to the one in iSpotNature.org, which provides a wide range of species to report. In contrast, other reporting services like Floracaching focus on one domain only, i.e. plants.

\subsubsection*{Factors of mobile UX}
An early listing of important dimensions affecting mobile UX, in general, was provided in \citep{arhippainen2003empirical}. These dimensions were: the user, the social factors, the cultural factors, the context of use, and the given product.  Along similar lines, \cite{Subramanya2007} divided factors contributing to a good user experience into three categories: device-related (i.e. hardware), communication-related (i.e. provide a feeling of face-to-face communication as much as possible), and application-related (i.e. UI related); \cite{wigelius2009dimensions} identified five dimensions affecting mobile UX: the social context, the spatial context, the temporal context, the infrastructural as well as the task contexts; and \cite{Ickin2012} identified seven factors that influence mobile UX in their study: application interface design, application performance, battery efficiency, phone features, application and connectivity cost, user routines, and user lifestyle. Concerning investigating UX, \cite{korhonen2010analyzing} proposed that mobile UX can be traced back to two features of the user's context: the triggering context (i.e. single contextual factor that changes the user's experience stream in a positive or negative direction), and the core experience (i.e. experience that was most memorable to the user during the interaction). The lessons learned about UI factors in this work intend to extend this body of knowledge with insight from UI element design and placement.

\subsubsection*{Form design on mobile devices}

\cite{Harms2015} investigated the use of long forms on smartphones. They tested four different designs and found scrolling to perform the worst of all possible methods, while the other three designs (tabs, menus, and collapsible fieldsets) worked equally well. This finding matches the framework of \cite{zhang2005challenges}, who stated that the cognitive load needed from the user should be minimized by avoiding long lists. One way to avoid scrolling in long forms is to structure the content in categories, each fitting the screen size. Besides, the framework from \citep{zhang2005challenges} proposes to use as little interaction as possible (which changes in categories would imply). Hence, the structuring of content on mobile phones should be designed thoughtfully. While \cite{couper2011placement} investigated the placement of buttons in long forms (and not their structuring), their statement that the design of forms can affect the user behavior speaks in favor of further investigating form design in general. These studies all examined the design of, and interaction with long forms. The question remains whether the findings are also applicable to short forms, and if scrolling in forms should be avoided in general.

\subsubsection*{Interactive maps}
\cite{degbelo2019interactive} compared the merits of form-based and map-based interaction for geodata creation tasks. They reported that the sweet spot of interactive maps (on desktop devices) seems not so much their impact on productivity, but rather their positive influence on the overall UX. Users reported that maps are more stimulating and attractive than forms for information creation tasks.  
Regarding the mobile context of interactive maps, \cite{burigat2011} investigated mobile maps and the use of three different approaches to visualizing the references to off-screen content of maps. In another study, \cite{burigat2008} found zoomable maps with an overview window (overview+detail approach) to be useful for map interaction. A third study by \cite{burigat2013} showed increased performance in task completion time with overview+detail layouts, which users could manipulate. The focus was on interfaces, which allow map manipulation through interactions with the overview and highlighting of objects. Finally, \cite{horbinski2020graphic} provided evidence that positioning of UI elements on mobile maps \textit{matters}. Users indeed expressed different preferences for the positioning of buttons for features such as geolocation, search, or routing. Though these studies provide valuable insights into the use of interactive maps on mobile devices, they do not cover an understanding of factors affecting mobile map user experience, which is the topic of this article.

\section{User study}
The goal was to investigate differences in users’ perception of the navigation modalities and sequences of the interactions. The interfaces were designed to investigate the influence of map positioning as well as different form designs on a smartphone. This study can be viewed, following \cite{Schmidt2009}, as a ‘follow-up study with a replication condition’ to \cite{Harms2015}'s work. The focus was on two questions. RQ1: what is the influence of different form designs on user experience and usability? The focus here was on short forms (i.e. up to 10 fields). RQ2: what is the influence of UI component sequences on user experience and usability?

\subsubsection*{Variables} Independent variables: The sequence of the UI components, and the design of the form UI were variables controlled by the different prototypes; dependent variables: usability, user experience, and task performance. Usability and user experience were measured using the System Usability Scale (SUS, \cite{brooke1995sus}) and the User Experience Questionnaire (UEQ-S, \cite{schrepp2017design}) respectively. Since it was not possible to integrate interaction logs into “Meine Umwelt”, the time to complete the tasks was measured using screen recordings of the smartphone used during the experiment for all prototypes.

\subsubsection*{Study design} 
\begin{figure*}
    \centering
    \frame{\includegraphics[scale=0.19]{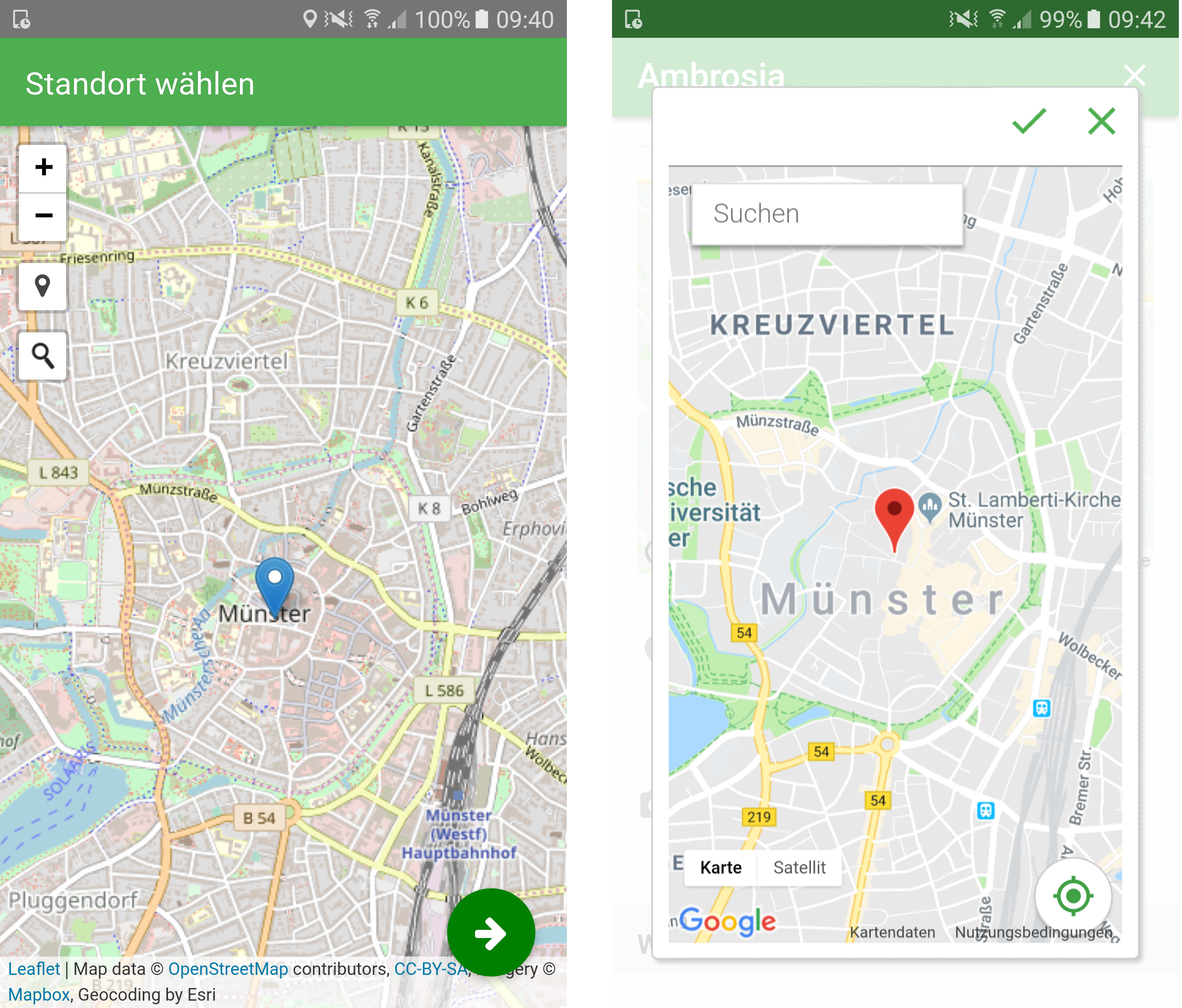}}
    \vspace{0.1cm}
   \frame{ \includegraphics[scale=0.19]{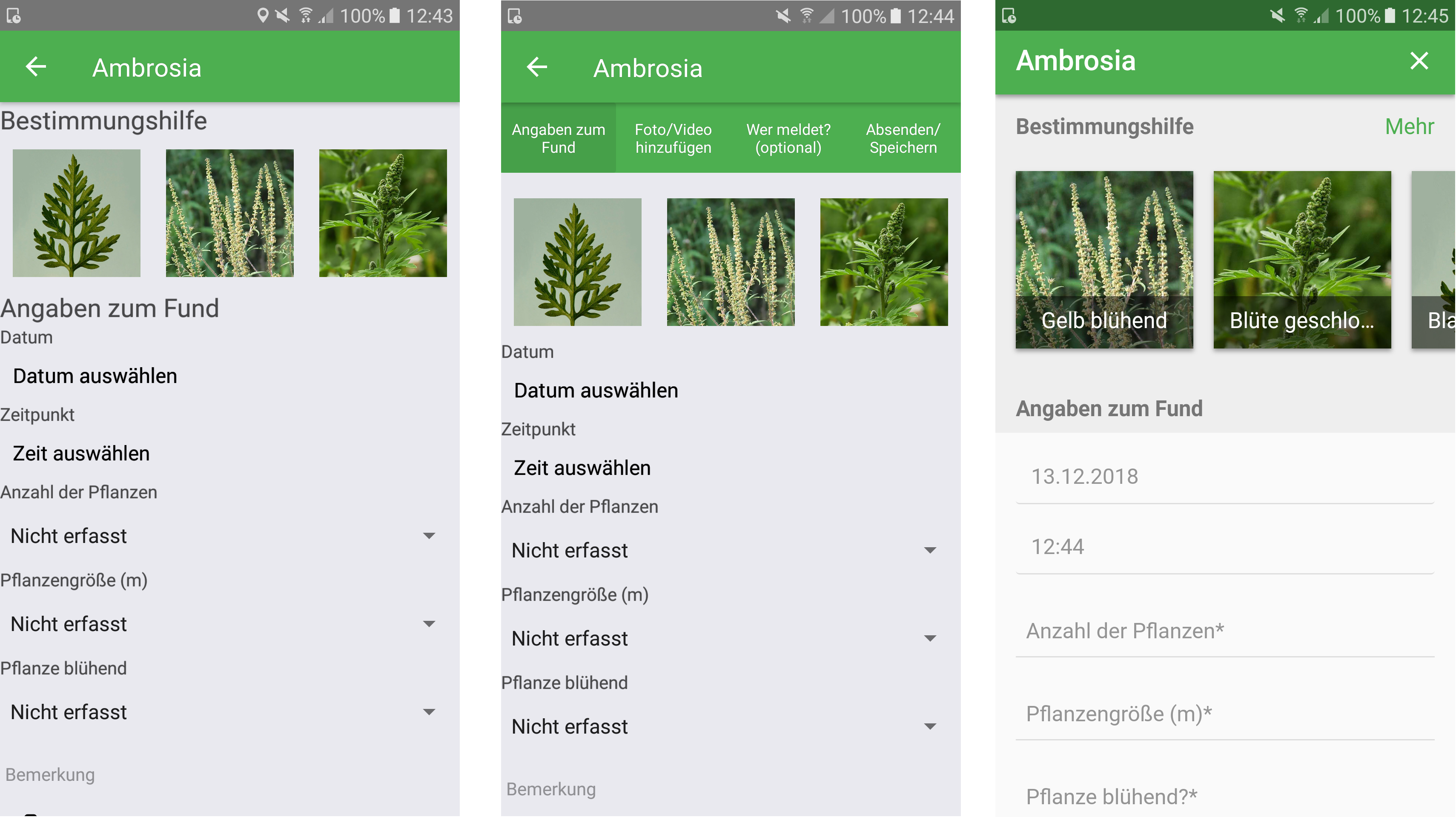}}
   \vspace{0.1cm}
     \frame{ \includegraphics[scale=0.19]{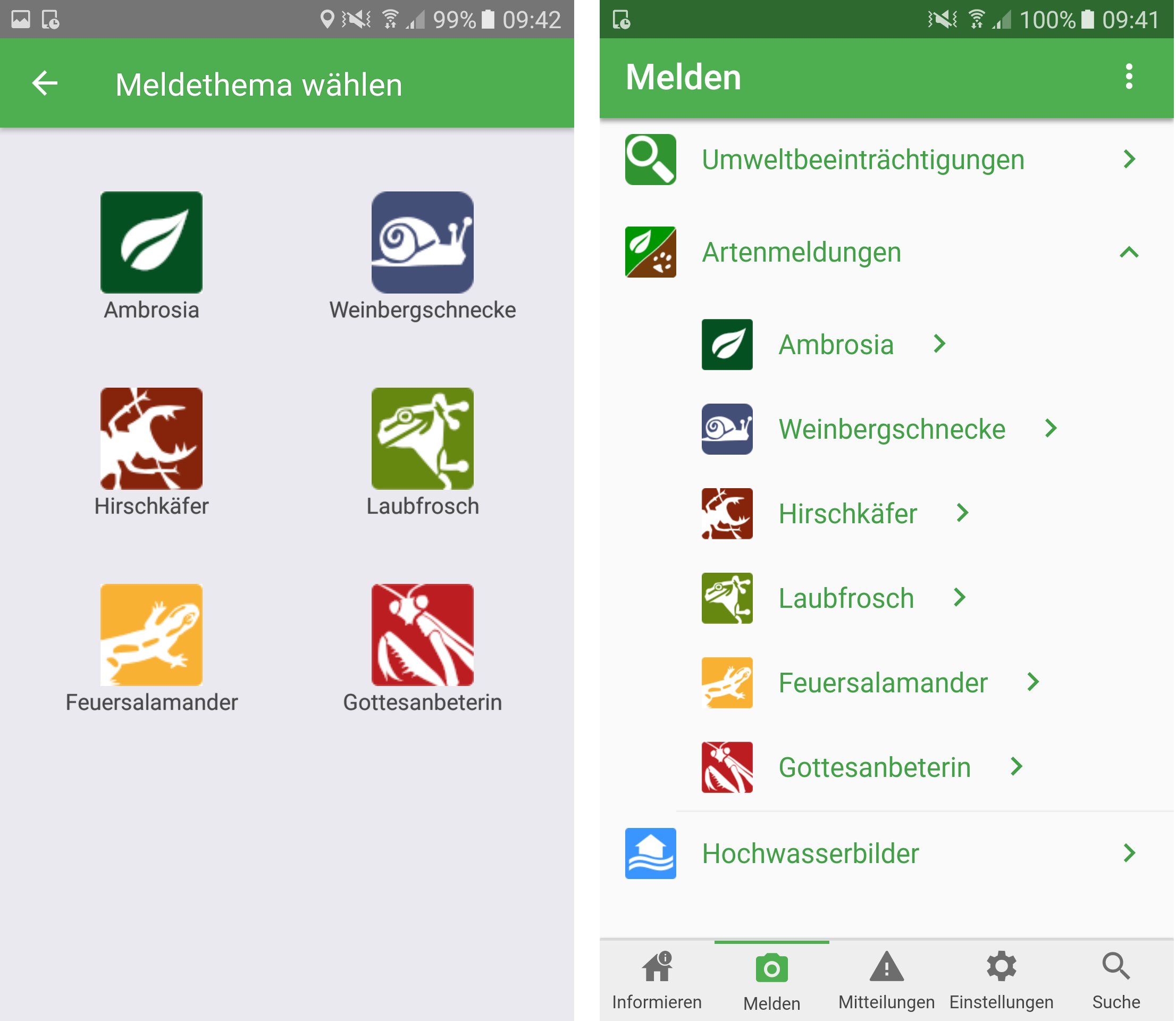}}
    \caption{Top: P1 \& P2 (map as landing screen) vs MU (map integrated into a form); Middle: P1 with scrollable form vs P2 with tabs to structure the form vs MU (Baseline); Bottom: Example species type in P1 \& P2 vs example species type in MU.}
    \label{fig:prototypes}
\end{figure*}

Besides the application “Meine Umwelt”, two prototypes were developed for the experiment (Figure \ref{fig:prototypes}). The prototypes were designed to separate the sequences of the UI elements, and the form designs. Therefore, three applications were used: Map + Selection + Form (scroll) [hereafter, Prototype 1]; Map + Selection + Form (tab) [hereafter, Prototype 2]; Selection + Form (scroll) + map [hereafter, Base condition] (“Meine Umwelt”). A within-subject design was used in the experiment. Each participant completed three ecological reporting tasks, with different levels of difficulty. The order of the tasks was counterbalanced using a Latin Square approach (see supplementary material).

\subsubsection*{Tasks} Each prototype was tested with three different tasks of equivalent properties and difficulty. Properties were the \textit{complexity of the task}, the \textit{complexity of the place}, and the \textit{point of time}. The \textit{complexity of the task} refers to different forms provided in the tasks. The tested condition of the “Base condition” provides six categories to report. These were all used in the tasks, while some of these had to be used multiple times to reach the number of nine tasks. The plant category “Ambrosia" was used in all task groups, because the complexity of this task was higher than for all other conditions. The form to report Ambrosia had three input fields more than the remaining categories, resulting in a more demanding task (as more information has to be remembered and filled in by the user). Each task group had one task with a \textit{point of time} requiring the user to go back in the calendar for several months, as opposed to the other dates used (e.g. “right now”, “yesterday”). Lastly, the \textit{complexity of the place} was higher for one task per group. This was used to make users interact with the map more than just searching for street names. The tasks were in German and are available in the supplementary material\footnote{See \url{https://doi.org/10.6084/m9.figshare.13174550}.}. The study was approved by the local ethics board and pilot tested with two participants. A result of the pilot study was a slight adjustment of the tasks, to improve their understandability. Also, a search bar was implemented for the prototypes to enable place searches on the map. The results from the two participants are not included in the analysis.

\subsubsection*{Participants} 18 participants (10 females and 8 males) were recruited by e-mail and word of mouth. Participants all spoke German because the study and app “Meine Umwelt” were in German. The participants included four landscape ecologists and biologists, who are involved in landscape and environmental preservation, along with other users, who might use the app more casually. This mix ensures a representative group to the user group of “Meine Umwelt”. The average age of the participants was 22.7 (SD: 2.7).

\section{Results} Table \ref{tab:results} presents the results of the study. In line with recommendations from previous work \cite{Dragicevic2016}, the whole analysis was done using interval estimates. Confidence intervals provide much richer information than p-values alone. A confidence interval that does not include zero indicates statistical significance; the tighter the interval, and the farther from zero, the stronger the evidence. The analysis was done using the BootES package \citep{Kirby2013} in R. The number of bootstrap resamples was N = 2000. Task performance (i.e. TimeOnTask) was similar across all three prototypes. 

Regarding usability, the Base condition “Meine Umwelt” had an average value of 59.7. The difference between the Base condition and Prototype 1 was significant (9.4, 95\%CI [1.5, 18.1]), and so was the difference between the Base condition and Prototype 2 (7.9, 95\%CI [0.7, 15.9]). In contrast, the difference between Prototype 1 and Prototype 2 was not significant with respect to usability ratings.

As regards user experience, the Base condition (hereafter MU) was rated with a pragmatic quality of 1.01 by the participants. The hedonic quality was lower at 0.38 and the overall rating was 0.70. Overall, users rated the user experience of the Prototype 1 (P1) significantly higher than the user experience of the Base condition (+0.57, 95\%CI [0.25, 0.97]). Prototype 2 (P2) collected also significantly higher ratings than the Base condition (+0.54, 95\%CI [0.14, 0.96]) for the overall user experience. The difference between MU and P1 was significant for the pragmatic quality (0.85, 95\%CI [0.2, 1.6]), and so was the difference between the MU and P2 for the pragmatic aspects (0.85, 95\%CI [0.01, 1.7]). The difference between P1 and P2 was not significant for the pragmatic quality. The differences between all prototypes were not significant for the hedonic quality. 

Regarding the size of the effects, Hedges' g was computed for all significant differences. This is in line with recommendations from previous work \citep[e.g.][]{Sauro2014a,Lakens2013} for studies with less than 20 participants. The values obtained were 0.52 (MU vs P1, usability); 0.46 (MU vs P2, usability); 0.71 (MU vs P1, user experience total); 0.56 (MU vs P2, user experience total); 0.50 (MU vs P1, user experience pragmatic); and 0.44 (MU vs P2, user experience pragmatic). These suggest a medium effect\footnote{Corresponding Cohen's d values were 0.54 (MU vs P1, usability); 0.48 (MU vs P2, usability); 0.74 (MU vs P1, user experience total); 0.59 (MU vs P2, user experience total); 0.53 (MU vs P1, user experience pragmatic); and 0.46 (MU vs P2, user experience pragmatic).}. 

Gender effects (male vs female) and background domain effects (landscape ecologist/biologist vs others) were tested. These were all non-significant on all dimensions (TimeOnTask, usability, and user experience). A learning effect was apparent in the data for task completion time. For 16 of 18 participants, the first app interaction took the most amount of time. Nevertheless, the effects were apparent for all applications, because of the Latin square design of the experiment and the resulting order of apps for each participant. Therefore, the learning effect did not influence the results of a single app. The main difference between the tested applications was the positioning of the map and the form designs. Both prototypes scored higher values for both usability and user experience. These prototypes were using the map as a landing screen for the application. The influence of form design was not apparent, as both prototypes scored similar values and performed with similar times. These results are now discussed in detail.

\begin{footnotesize}
\begin{table*}
\centering
\begin{tabular}{|l|l|l|l|l|l|l|}
\hline
 & MU & 95\% CI & P1 & 95\% CI & P2 & 95\% CI \\ \hline
\textbf{Time (seconds)} & 318 & {[}298, 339{]} & 306 & {[}279, 335{]} & 323 & {[}287, 379{]} \\ \hline
Time (Diff, MU) & \multicolumn{2}{l|}{\cellcolor[HTML]{C0C0C0}} & -13 & {[}-42, 23{]} & 4 & {[}-38, 56{]} \\ \hline
Time (Diff, P2) & \multicolumn{2}{l|}{\cellcolor[HTML]{C0C0C0}} & -17 & {[}-84, 31{]} & \multicolumn{2}{l|}{\cellcolor[HTML]{C0C0C0}} \\ \hline
\textbf{SUS Score} & 59.7 & {[}52.1, 66.0{]} & 69.2 & {[}65.3, 72.4{]} & 67.6 & {[}63.2, 70.8{]} \\ \hline
SUS (Diff, MU) & \multicolumn{2}{l|}{\cellcolor[HTML]{C0C0C0}} & \cellcolor[HTML]{38FFF8}9.4 & \cellcolor[HTML]{38FFF8}{[}1.5, 18.1{]} & \cellcolor[HTML]{38FFF8}7.9 & \cellcolor[HTML]{38FFF8}{[}0.69, 15.83{]} \\ \hline
SUS (Diff, P2) & \multicolumn{2}{l|}{\cellcolor[HTML]{C0C0C0}} & 1.53 & {[}-3.3, 5.0{]} & \multicolumn{2}{l|}{\cellcolor[HTML]{C0C0C0}} \\ \hline
\textbf{UEQ (prag.)} & 1.01 & {[}0.32, 1.65{]} & 1.86 & {[}1.57, 2.10{]} & 1.86 & {[}1.38, 2.18{]} \\ \hline
UEQ p. (Diff, MU) & \multicolumn{2}{l|}{\cellcolor[HTML]{C0C0C0}} & \cellcolor[HTML]{38FFF8}0.85 & \cellcolor[HTML]{38FFF8}{[}0.22, 1.63{]} & \cellcolor[HTML]{38FFF8}0.85 & \cellcolor[HTML]{38FFF8}{[}0.01, 1.69{]} \\ \hline
UEQ p. (Diff, P2) & \multicolumn{2}{l|}{\cellcolor[HTML]{C0C0C0}} & 0 & {[}-0.38, 0.44{]} & \multicolumn{2}{l|}{\cellcolor[HTML]{C0C0C0}} \\ \hline
\textbf{UEQ (hed.)} & 0.38 & {[}-0.03, 0.82{]} & 0.64 & {[}0.13, 1.14{]} & 0.61 & {[}0.17, 1.03{]} \\ \hline
UEQ h. (Diff, MU) & \multicolumn{2}{l|}{\cellcolor[HTML]{C0C0C0}} & 0.26 & {[}-0.05, 0.52{]} & 0.24 & {[}-0.01, 0.44{]} \\ \hline
UEQ h. (Diff, P2) & \multicolumn{2}{l|}{\cellcolor[HTML]{C0C0C0}} & 0.23 & {[}-0.22, 0.25{]} & \multicolumn{2}{l|}{\cellcolor[HTML]{C0C0C0}} \\ \hline
\textbf{UEQ (total)} & 0.70 & {[}0.23, 1.17{]} & 1.27 & {[}0.98, 1.51{]} & 1.24 & {[}0.97, 1.60{]} \\ \hline
UEQ t. (Diff, MU) & \multicolumn{2}{l|}{\cellcolor[HTML]{C0C0C0}} & \cellcolor[HTML]{38FFF8}0.57 & \cellcolor[HTML]{38FFF8}{[}0.25, 0.97{]} & \cellcolor[HTML]{38FFF8}0.54 & \cellcolor[HTML]{38FFF8}{[}0.14, 0.96{]} \\ \hline
UEQ t. (Diff, P2) & \multicolumn{2}{l|}{\cellcolor[HTML]{C0C0C0}} & 0.03 & {[}-0.19, 0.30{]} & \multicolumn{2}{l|}{\cellcolor[HTML]{C0C0C0}} \\ \hline
\end{tabular}
\caption{Results. Average values per prototype are reported first, followed by the average of the differences between the prototypes. Cells highlighted in blue indicate significant values (i.e. confidence intervals that do not include zero).}
\label{tab:results}
\end{table*}
\end{footnotesize}

\section{Discussion}
This section touches upon five topics: interpretation of the task performance results, the effect of form design on UX (RQ1), the effect of the sequence of UI elements on UX (RQ2), implications of the results for theory and design, and the limitations of the study. 

\subsubsection*{Functional similarity of the apps} The results of the user study have shown similar results for the task performance of the tested applications. All three of them need approximately the same time to complete all tasks (Table \ref{tab:results}). This indicates that the applications are all equally suited for the task at hand, namely the reporting of ecological data. There were no clear indications for an app hindering users in effectively performing the provided tasks. A goal of designing the two additional prototypes was to provide a similar design, to be able to compare outcomes. With all applications performing on a similar level regarding the time required to solve the tasks, this goal seems to have been achieved.

\subsubsection*{Effect of form design} Considering the first research question, the results have shown no significant difference between scrollable forms and tab-structured forms. Put differently, the sequences ``Map + Selection + Form (scroll)'' [P1] and ``Map + Selection + Form (tab)'' [P2] can be considered equivalent with regard to task completion, usability and UX. The main implication is that the findings of \cite{Harms2015}, who found scrolling to be worse (as to usability) for the navigation in long forms, cannot be generalized to all sizes of forms. The usability of forms of a size up to 10 input fields is not significantly improved by a tab-based design. Thus, form designs do not noticeably influence usability and user experience in all cases.  

\subsubsection*{Effect of sequence of UI elements} Significant differences between the Base condition (MU) and Prototype 1 were found for the user experience ratings. The differing variable between the two applications was the positioning of the map. Looking closely at the dimensions of user experience, hedonic UX was not significantly impacted, while pragmatic UX was rated differently by the participants. Likewise, usability ratings were much higher for Prototype 1. This answers the second research question: positioning the interactive map before the form in the sequence of interactions results in better user experience and usability. Given that perceived usability is strongly associated with pragmatic user experience (see e.g. \cite{diefenback2014hedonic,Pettersson2018}), the convergence of usability and UEQ ratings during the experiment increases our confidence that \textit{pragmatic} user experience has been positively affected by the change in the positioning of the map component. In addition, given that the sequences ``Map + Selection + Form (scroll)'' [P1] and ``Map + Selection + Form (tab)'' [P2] can be considered equivalent (see above), the fact that similar observations are made for Prototype 2 also provide confirmatory evidence for the result. There may be several reasons for this, and these are now discussed using two theories: `cognitive fit theory' \citep{Vessey1991} and `memory-based theory of sequence effects' \citep{Cockburn2017}. 

One reason could be that the sequence ‘location reporting =$>$ species type selection =$>$ specie reporting’ (sequence 1) provides a better cognitive fit than the sequence ‘species type selection =$>$ specie reporting =$>$ location reporting’ (sequence 2). Cognitive fit theory \cite{Vessey1991} posits that problem-solving efficiency and effectiveness increases when problem-solving aids (in this context UI elements) support strategies required to perform a task. In the case of sequence 2, the map popping up to enter the location \textit{interrupts} the interaction of filling a form (see Figure \ref{fig:prototypes}, left). In contrast, sequence 1 provides a full-screen map interface to enter the location and additionally encapsulates the location input. Hence, users are done with entering location data, when they start to provide additional data (e.g. species type, date, height/number of plants, see Figure \ref{fig:prototypes}, middle) using forms. Thus, sequence 1 offers a `cleaner' task separation and hence a better cognitive fit. It is worth mentioning that cognitive fit in the strictest sense only predicts improved efficiency and effectiveness for some conditions. These were not observed (see Table \ref{tab:results}). Yet, if the scope of cognitive fit theory is broadened to non-instrumental aspects of interaction, the explanation above is plausible. A single experiment may not justify broadening the scope of the theory, but based on the results, the following provisional postulate can be formulated: user experience and/or performance increase when problem-solving aids support strategies required to perform a task.

Another reason for differences in the scores might be that both SUS and UEQ measure the perceived usability and UX respectively. Therefore, the first impression of the map as the landing screen for an application might simply be \textit{perceived} more positively. This explanation would concur with \cite{Lindgaard2006}, who found that users form their judgment about the visual appeal of web pages during the first 50 milliseconds of their interaction. That finding was confirmed in \citep{Tractinsky2006}, who also provided evidence that first impressions about the attractiveness of web pages are not only formed quickly, but lasting. Since users find maps to be more stimulating and attractive than forms for information provision (see \cite{degbelo2019interactive}), it is plausible that their first impression seeing the map first was much more positive, and lasted till the end of the interaction. Besides, given that UX was measured after the experiment, a memory-based view of UX could be valuable at this point. According to memory-based theories of UX \citep{Cockburn2017}, three factors influence people’s memory of experiences: primacy (i.e. over-weighted influence of the initial moments of an experience), recency (i.e. over-weighted influence of the terminating moments of the experience), and peak-end (i.e. over-weighted influence of the most intense moments of the experience). Since the map was present at the beginning in sequence 1 and at the end in sequence 2, producing in one case a positive primacy effect, and in the other a positive recency effect, the results suggest that there may be cases where primacy effects with a UI element weight stronger than recency effects with that element. The explanations provided in this and the previous paragraph are arguably tentative, but are also useful hypotheses to corroborate in follow-up experiments.

\subsubsection*{Implications for theory and design} ``Meine Umwelt'' is an incident reporting app, and as such, the lessons learned apply to incident reporting apps more broadly. As to theory, \cite{Winckler2013} provided a comprehensive model for tasks related to incident reporting. In their model, they suggested that sequence does not matter for the sub-tasks: `describe the incident', `locate the incident', and `inform time for the incident'. The results above suggest that make the \textit{scope} of their theoretical model more precise. Sequence does not matter from the time completion point of view but does matter from the pragmatic UX point of view. As to design, \cite{Norman2009} calling designers to design for memorable experiences, formulated these rules of thumb: ``What is the most important part of an experience? Psychologists emphasize what they call the primacy and recency effects, with recency being the most important. In other words, what is most important? The ending. What is most important after that? The start. So make sure the beginning and the end are wonderful''. The end may not always be more important than the start according to the results above.

\subsubsection*{Limitations}
Though the prototypes were designed as closely as possible to the baseline application “Meine Umwelt”, some minor differences can still be detected when comparing the applications side by side (e.g. captions in the pictures, or the choice of the map tile provider, see Figure \ref{fig:prototypes}, left). These differences could not be removed, and stem from the different technologies used for the app development. “Meine Umwelt” uses an API key for Google Maps (and thus Google Maps as a base map) while the prototypes implemented used Open Street Map as a base map. Besides, while “Meine Umwelt” used the Cordova framework, React Native was used to develop the prototypes. The native components provided by React Native have some slightly different properties than those of Cordova (e.g. the placeholders for $<$input$>$ tags displayed in a slightly different way, see Figure \ref{fig:prototypes}, middle). These differences (i.e. map[googlemaps] vs map [openstreetmap] or form [cordova] vs form [reactnative]) are minor nonetheless and are mentioned here for the sake of completeness only\footnote{There is no recent study known to the authors reporting on a comparison between the user experience - on mobile devices - of Google Maps vs. OpenStreetMaps, and the user experience of Cordova Apps vs. React Native apps. Nonetheless, a study by \cite{Schnur2018} reported that the perceived user complexity of Google Maps was consistently lower than that of OpenStreetMaps for several levels of details. \cite{Tuch2009}, in the context of websites, reported an inverse-linear relationship between visual complexity and pleasure. That is, start pages with low visual complexity were rated by users as more pleasurable. A replication of several studies \citep{Miniukovich2020} confirmed that inverse-linear relationships for websites. \cite{Miniukovich2014} also observed a negative correlation between visual complexity and aesthetics for mobile apps. Putting these findings together suggests - if the difference map [googlemaps] vs. map [openstreetmap] was important - that P1 \& P2 (OpenStreetMaps, Figure \ref{fig:prototypes}, top left) would have obtained lower user experience ratings than MU (Google Maps, Figure \ref{fig:prototypes}, top right). The opposite was observed during the study.}.

\section{Conclusion}
This article has investigated the effect of sequence of UI elements and type of forms within a mobile application for reporting ecological data. A user study has shown significant preferences for the map as a starting element, instead of the map as an ending element. Besides, a tab-based, structured design was tested against a scrollable view. Results have shown no significant difference between these designs in short forms. With respect to earlier research \citep{Harms2015}, it can be concluded, that scrolling does not perform worse for all sizes of form length. In short, the sequence of user interface elements on mobile devices matters, and the type of form design matters, depending on the length of the forms. Designers should keep in mind both (besides button placement identified in previous work \cite{horbinski2020graphic}) while building their next incident reporting app. Future work can replicate this study, for instance, factoring in data about the complexity of the base maps, and collecting qualitative data about what users dis/liked. Additionally, future studies can further investigate why perceived user experience and usability are better for some sequences than for others, based on a revised version of the cognitive fit theory, and memory-based theories of UX.  

\subsubsection*{Acknowledgments} We thank xdot GmbH for their support and for sharing the code of “Meine Umwelt”.

\bibliographystyle{apalike}
{\small
\bibliography{example}}

\end{document}